%% file: eprint.tex
\documentclass[10pt]{article}
\usepackage{graphicx}

\def\Title#1{\begin{center} {\Large #1 } \end{center}}
\def\Author#1{\begin{center}{ \sc #1} \end{center}}
\def\Address#1{\begin{center}{ \it #1} \end{center}}

\newcommand\pubblock{\rightline{\begin{tabular}{l} Proceedings of the Second Annual LHCP\\ \pubnumber\\
         \pubdate  \end{tabular}}}

\newenvironment{Abstract}{\begin{quotation} \begin{center} 
             \large ABSTRACT \end{center}\bigskip 
      \begin{center}\begin{large}}{\end{large}\end{center} \end{quotation}}

\newenvironment{Presented}{\begin{quotation} \begin{center} 
             PRESENTED AT\end{center}\bigskip 
      \begin{center}\begin{large}}{\end{large}\end{center} \end{quotation}}

\def\Acknowledgements{\bigskip  \bigskip \begin{center} \begin{large}
             \bf ACKNOWLEDGEMENTS \end{large}\end{center}}

\input econfmacros.tex

\usepackage{color}

\newcommand\pubnumber{ CMS-CR-2014/193 }

\newcommand\pubdate{\today}

\def\affiliation{
On behalf of the CMS Collaboration, \\
Department of Physics \& Astrophysics \\
University of Delhi, New Delhi 110007, INDIA }

\begin{document}

\large
\begin{titlepage}
\pubblock

\vfill
\Title{  Measurement of Electroweak Vector Boson Pair Production in \textit{pp} Collision with the CMS
Detector at LHC  }
\vfill

\Author{ AJAY KUMAR  }
\Address{\affiliation}
\vfill
\begin{Abstract}

We present an overview of measurements of electroweak vector boson pair production and anomalous Triple Gauge Couplings (aTGC),
with semileptonic and fully leptonic final states. The data analyzed were taken at center of mass energy of 7 \& 8 TeV by the CMS detector
at the Large Hadron Collider. The cross-section measurements are important because they are test of the Standard Model predictions,
while these processes serve as background for Higgs searches and various other processes.

\end{Abstract}
\vfill

\begin{Presented}
The Second Annual Conference\\
 on Large Hadron Collider Physics \\
Columbia University, New York, U.S.A \\ 
June 2-7, 2014
\end{Presented}
\vfill
\end{titlepage}
\def\thefootnote{\fnsymbol{footnote}}
\setcounter{footnote}{0}
%

\normalsize 

\section{Introduction}
Electoweak Vector Boson pair production provides an important test of  the Standard Model (SM) of particle physics. These processes are
backgrounds to important searches such as Higgs Boson, Supersymmetric particles etc. and their understanding are necessary for exploring $VV$
scattering. Further, any deviation in the production cross-sections or in kinematics from the SM predictions can provide a possible hint to
new physics. The experimental results reviewed here use $pp$ collision at the Large Hadron Collider (LHC) at a center-of-mass energy
$\sqrt{s}$ = 7 and 8 TeV, with an integrated luminosity up to about 19.6 fb$^{-1}$. 
\section{Cross Section Measurements}
\subsection{$ZZ$ in four leptons}
The $ZZ$ production cross section has been measured in the following final states, $ZZ \rightarrow \ell\ell\ell'\ell'$, 
where $\ell$ = e, $\mu$, and $\ell'$ = e,$\mu$ and $\tau$~\cite{Khachatryan:2014dia}
and $ZZ \rightarrow 2\ell2\nu$~\cite{CMS:2013qez,Chatrchyan:2012sga}.
For $ZZ \rightarrow \ell\ell\ell'\ell'$ signal is characterized by four charged leptons. Events are selected to get mutually exclusive set of signal candidates in the $ZZ \rightarrow \ell\ell\ell\ell$, and $ZZ \rightarrow \ell\ell\tau\tau$ channels. $Z$ candidate is formed using a pair of well-identified and isolated leptons of the same flavor and opposite charge with invariant mass consistent with the nominal mass of $Z$ boson. The major background contributions arise from the production of $Z$ and $WZ$ in association with jets and from $t\bar{t}$. In all these cases, a jet or a non-isolated lepton is misidentified as a lepton. We estimate these backgrounds in data-driven way using appropriate control region by relaxing isolation criterion.
In $ZZ \rightarrow 2\ell2\nu$,the signal consists of two $Z$ bosons such that one decays into a pair of opposite-charged leptons ($e$ or $\mu$), and the other to two neutrinos ($\nu$) that escape direct detection~\cite{CMS:2013qez,Chatrchyan:2012sga}. The final state is thus characterized by a pair of opposite-charge, isolated electrons or muons, with invariant mass close to nominal mass of $Z$ boson; no additional leptons, other than the muon or electron pair from the $Z$ decay; and large missing transverse energy (MET). $Z$ boson mass criterion suppresses all backgrounds which do not have real $Z$ boson while lepton veto suppresses $WZ$ background. Given that the $ZZ$ pair is produced in the collision of two hadrons, extra jets may be found in the event. In order to minimize backgrounds coming from top quarks,
events are vetoed if a b-tagged jet 
is found. Furthermore, the MET based variables are used to suppress Drell-Yan (DY) background.
Data-driven method is used to estimate DY background using $\gamma$+jets control sample. We also use
another data-driven method to estimate background events which does not involve $Z$ boson such as
$WW$ and top quark production. 
\subsection{$WZ$ in 3$\ell\nu$}
The signal signature is characterized by a pair of same-flavor, opposite-charge,
isolated leptons with an invariant mass consistent with the $Z$ boson, together with
a third isolated lepton and a significant amount of MET associated
with the escaping neutrino~\cite{CMS:2013qea}.
%
Backgrounds are grouped as: non-peaking background such as $t\bar{t}$, QCD multijet and $W$+jets
production; $Z$+fake lepton background; and $Z$+prompt lepton background. Z+fake lepton backgrounds are the most important for
the $WZ \to 3\ell\nu$ process; they include fake or non-isolated leptons from
jets (including heavy quark jets) or photons. $Z$+prompt lepton backgrounds
originate primarily from the $ZZ \to 4\ell$ process, where one of the four
leptons is lost; it is irreducible but small due to relatively small production
cross section. The contributions from other backgrounds are negligibly small.
\subsection{$WW$ in 2$\ell$2$\nu$}
The signal is characterized by two opposite charged isolated leptons (electrons or muons) accompanied by significant amount of
MET~\cite{Chatrchyan:2013oev,Chatrchyan:2013yaa}.
Background processes in this analysis include \textit{W}+jets, \textit{W}$\gamma$, and QCD multijet events where at
least one of the jets is misidentified as a lepton, top production
($t\bar{t}$ and \textit{tW}), DY, and diboson
production (\textit{WZ} and \textit{ZZ}). Top contribution is suppressed by applying additional jet veto.
DY acts as a background in two cases: first, when poorly reconstructed leptons or jets are
mismeasured as MET, which is suppressed by requiring \textit{projected $E_{T}^{miss}$} above certain threshold.
Second, when the \textit{Z} boson recoils against a jet. In this case the angle in the transverse plane between the dilepton system and
the most energetic jet with transverse energy ($E_{T}$) above 15 GeV is required to be smaller than 165 degrees.
This selection is applied only in the \textit{e}$^{+}$\textit{e}$^{-}$ and \textit{$\mu$}$^{+}$\textit{$\mu$}$^{-}$
final states.
Further DY background in the
\textit{e}$^{+}$\textit{e}$^{-}$ and \textit{$\mu$}$^{+}$\textit{$\mu$}$^{-}$ final states is reduced by requiring dilepton mass within
15GeV of the \textit{Z} mass.
Events with dilepton masses below 12 GeV are also rejected
to suppress contributions from low-mass resonances. The same
requirement is also applied in the \textit{e}$^{\pm}$\textit{$\mu$}$^{\mp}$ final state. 
Diboson processes, such as \textit{WZ} and \textit{ZZ}
production, are suppressed by third lepton veto.
\textit{W$\gamma$} production, in which photon converts, is suppressed
by rejecting electrons consistent with a photon conversion.
A combination of data-driven method and
Monte Carlo (MC) simulation studies are used to estimate background contributions.
\subsection{$WW+WZ$ in $\ell\nu$$jj$}
Events are selected with one well-identified and isolated lepton (muon or electron), large MET,
and exactly two high-$p_{T}$ jets\cite{Chatrchyan:2012bd}.
The advantage of reconstructing \textit{WW+WZ} in semileptonic final state over the purely leptonic process is that the \textit{W} and \textit{Z}
bosons have large branching ratio to quarks decay than lepton decays. However, this is partially nullified due to
presence of huge \textit{W}+jets background having large cross section. Backgrounds to this process arise from \textit{W/Z}+jets, QCD multijet, $t\bar{t}$ and single top. Top is suppressed by additional jet veto.
To reduce the background from processes that do not contain $W \rightarrow \ell\nu$ decays, transverse
mass of the \textit{W} candidate is required above 30(50) GeV in $\mu$ ($\ell$) data.
To suppress DY and electroweak diboson purely-leptonic decay processes, we use additional lepton veto.
QCD multijet background is estimated in a data-driven way from two component fit to data on MET distribution, which determines the
corresponding fraction in data.
The signal is extracted by an unbinned maximum likelihood fit over dijet mass distribution in the mass range 40-150GeV
keeping diboson contribution normalization as free parameter. The normalization of other backgrounds are allowed to vary
within Gaussian constraints around their central values. The central values of these backgrounds are taken from next-to-leading or 
higher order theoretical calculations.
\subsection{$VZ$ in $VZ \rightarrow Vb\bar{b}$}
The $VZ$, where $V =W/Z$ production cross-section has been measured
in the $VZ \rightarrow Vb\bar{b}$ decay mode with, the subsequent decay mode of $V$ being $Z \rightarrow \nu\bar{\nu}$,
$W \rightarrow \ell\nu$ and $Z \rightarrow \ell\ell$ ($\ell = e, \mu$)~\cite{Chatrchyan:2014aqa}.
The event selection is based on the reconstruction of the vector bosons in their leptonic decay modes and of the $Z$ boson decay
into two b-tagged jets. Dominant backgrounds to $VZ$ production originate from $V$+heavy flavor (HF) jets, $V$+LF light flavor (LF) jets,
$t\bar{t}$, single top, multi-jet QCD and Higgs production.
In general b-tagging is used to reduce LF components, additional jet activity veto is used to reduce $t\bar{t}$ and
single top events.
The cross sections are extracted simultaneously for $WZ$ and $ZZ$ production
for the inclusive phase space and in a fiducial region
for a $V$ transverse momentum above 100 GeV with the $Z$ bosons
produced in the mass region $60 < M_{Z} < 120$ GeV.
\begin{figure}[htb]
\centering
\includegraphics[height=2in]{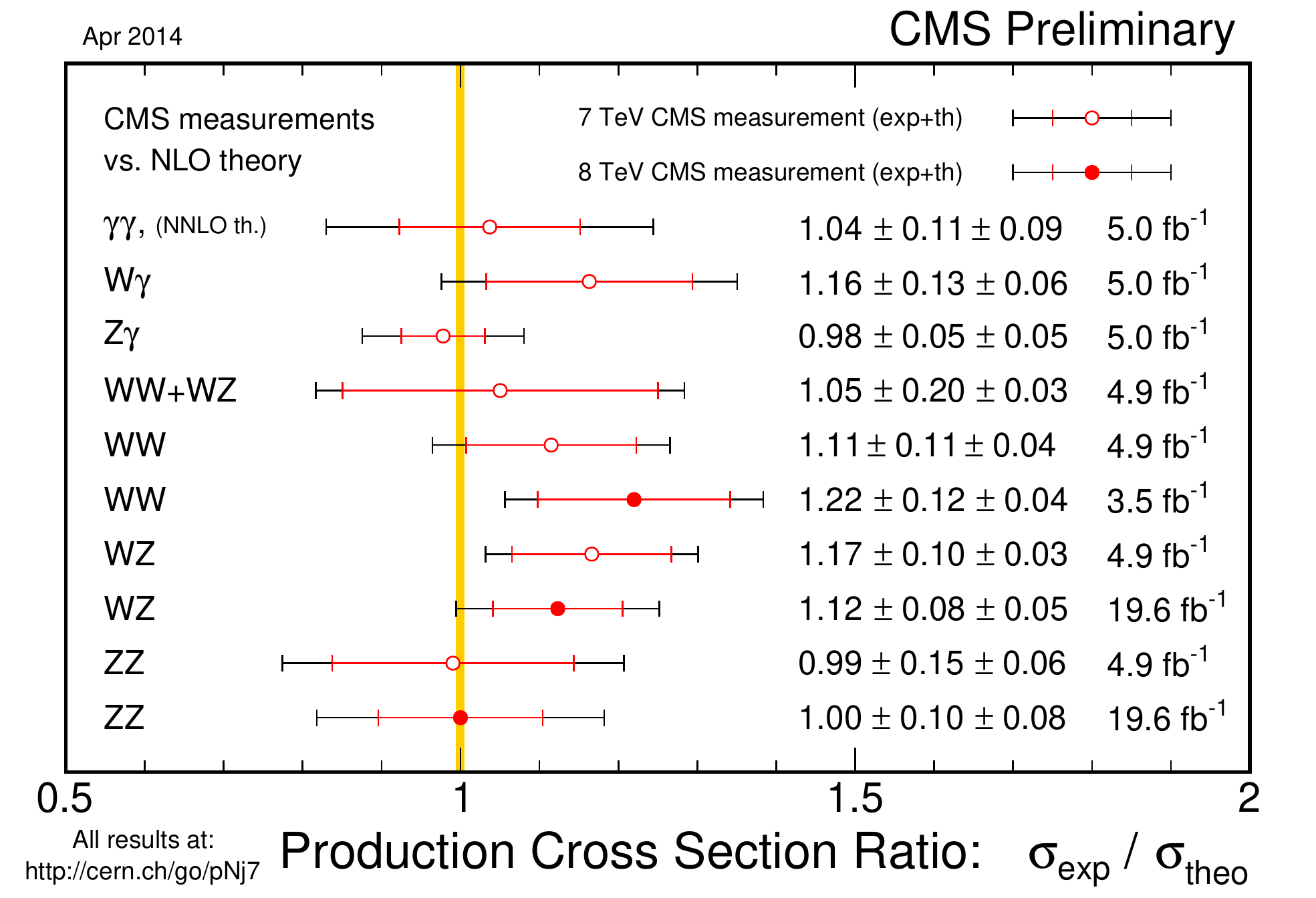}
\caption{ Relative Production cross-section for various Electroweak boson pairs w.r.t. SM}
\label{fig:figure1}
\end{figure}
\begin{table}[t]
\begin{center}
\begin{tabular}{l|ccc}  
Process & Measured ($\sigma$ in \rm{pb}) & Predicted($\sigma$ in \rm{pb}) & Lumi. in $fb^{-1}$ \\ \hline
$ZZ$ in $\ell\ell\ell'\ell'$ &$7.7^{+0.5}_{-0.5}~(\rm{stat.}) ^{+0.5} _{-0.4}~(\rm{syst.}) \pm 0.4~ (\rm{theo.)} \pm 0.3~(\rm{lumi.})$ &
${\rm 7.7 \pm 0.6}$  & 19.6 at 8 TeV  \\
$ZZ$ in 2$\ell$2$\nu$ &$6.8_{-0.8}^{+0.8}\,(\mathrm{stat.})\,_{-1.4}^{+1.8}\,(\mathrm{syst.})\,\pm 0.3\,(\mathrm{lumi.})$ &$7.92^{+4.7\%}_{-3.0\%}$ &     19.6 at 8 TeV  \\
$WZ$ in 3$\ell\nu$  &$24.61 \pm 0.76~({\rm stat.}) \pm 1.13~({\rm syst.}) \pm 1.08~({\rm lumi.})$ & $21.91_{-0.88}^{+1.17}$ & 19.6 at 8 TeV  \\
$WW$ in in 2$\ell$2$\nu$  &$69.9  \pm 2.8~({\rm stat.})  \pm 5.6~({\rm syst.}) \pm 3.1~({\rm lumi.})$ & $57.3 \left(^{+2.4}_{-1.6}\right)$ & 3.54 at 8 TeV    \\
$WW+WZ$ in $\ell\nu$\textit{jj} &$68.9 \pm 8.7~({\rm stat.}) \pm 9.7~({\rm syst.}) \pm 1.5~({\rm lumi.})$ &$ 65.6 \pm 2.2$ & 5.0 at 7 TeV   \\
$WZ$ in $Wb\bar{b}$ &$30.7 \pm 9.3~({\rm stat.}) \pm 7.1~({\rm syst.}) \pm 4.1~(\rm{theo.)} \pm 1.0~({\rm lumi.})$ &$ 22.3 \pm 1.1$ & 18.9 at 8 TeV   \\
$ZZ$ in $Zb\bar{b}$ &$6.5 \pm 1.7~({\rm stat.}) \pm 1.0~({\rm syst.}) \pm 0.9~(\rm{theo.)} \pm 0.2~({\rm lumi.})$ &$ 7.7 \pm 0.4$ & 18.9 at 8 TeV   \\ \hline
\end{tabular}
\caption{ Observed and expected cross-section. }
\label{tab:table1}
\end{center}
\end{table}
\section{Limits on Anomalous Triple Gauge Couplings (aTGC)}
Many di-boson processes contain TGC vertices or are sensitive to aTGCs that are not allowed in the SM at tree level such as 
$ZZ$$\gamma$, $ZZZ$ etc.
The strategy for these analyzes is to consider all possible extensions to the SM Lagrangian that preserve gauge invariance and CP
symmetry and set limits on possible enhancements of these terms. In general, enhanced TGCs should result in excess of events in the 
boson high $P_{T}$ tails and at high mass. We therefore investigate these distributions to set limits. Analysis reviewed above 
gives limit on anomalous  $ZZZ$ and $ZZ$$\gamma$ couplings based on the invariant mass distribution of the four-lepton system at
8 TeV~\cite{Khachatryan:2014dia,Hagiwara:1986vm}.
The measured couplings are $-0.004<f_4^{Z,\gamma}<0.004$, and $-0.005<f_5^{Z,\gamma}<0.005$
at 95\% CL, which are in good agreement with the SM predictions. 
%
\section{Summary}
Figure \ref{fig:figure1} shows the relative production cross section for various electroweak boson pairs and Table \ref{tab:table1} lists the values of the observed cross sections together with the expectation from the SM. The measured results are consistent with the SM predictions, within uncertainties. Also, no evidence for anomalous coupling in any of above processes are seen. 
%
\Acknowledgements
I am thankful to Kalanand Mishra(Fermilab) and Ilya Osipenkov(TAMU) for fruitful discussions. I also would like to thank 
LPC, Fermilab and Delhi University for financial support.

\end{document}

%% file: econfmacros.tex
\def\beq{\begin{equation}}
\def\eeq#1{\label{#1}\end{equation}}
\def\eeqn{\end{equation}}

\def\beqa{\begin{eqnarray}}
\def\eeqa#1{\label{#1}\end{eqnarray}}
\def\eeqan{\end{eqnarray}}

\let\bar=\overbar

\def\Dslash{\not{\hbox{\kern-4pt $D$}}}
\def\dslash{\not{\hbox{\kern-2pt $\del$}}}

\def\msb{{\bar{\ssstyle M \kern -1pt S}}}

